\documentclass[11pt]{article}
\usepackage{latexsym}
\usepackage{tabularx,booktabs,multirow,delarray,array}
\usepackage{graphicx,amssymb,amsmath,amssymb}
\usepackage[vlined, ruled]{algorithm2e}

\def\pra{PRA}
\def\sangle{\angle^{\uparrow}}

\def\lemmaspace{\vspace*{0in}}
\def\sectionspace{\vspace*{0in}}

\begin{document}

\baselineskip=14.0pt

\title{
\vspace*{-0.55in} An Improved Algorithm for Reconstructing a Simple
Polygon from the Visibility Angles\thanks{This research was
supported in part by NSF under Grant CCF-0916606.}}

\author{
Danny Z. Chen\thanks{Department of Computer Science and Engineering,
University of Notre Dame, Notre Dame, IN 46556, USA.
E-mail: {\tt \{dchen, hwang6\}@nd.edu}.
}
\hspace*{0.2in} Haitao Wang\footnotemark[2] \thanks{Corresponding
author.
}}

\date{}

\maketitle

\thispagestyle{empty}

\newtheorem{lemma}{Lemma}
\newtheorem{theorem}{Theorem}
\newtheorem{corollary}{Corollary}
\newtheorem{fact}{Fact}
\newtheorem{definition}{Definition}
\newtheorem{observation}{Observation}
\newtheorem{condition}{Condition}
\newtheorem{property}{Property}
\newtheorem{claim}{Claim}
\newenvironment{proof}{\noindent {\textbf{Proof:}}\rm}{\hfill $\Box$
\rm}

\pagestyle{plain}
\pagenumbering{arabic}
\setcounter{page}{1}

\vspace*{-0.2in}
\begin{abstract}
In this paper, we study the following problem of reconstructing a
simple polygon: Given a cyclically 
ordered vertex sequence of an unknown simple polygon $P$ of $n$ vertices and, for each vertex
$v$ of $P$, the sequence of angles defined by all the visible vertices of $v$ in $P$,
reconstruct the polygon $P$ (up to
similarity). An $O(n^3\log n)$
time algorithm has been proposed for this problem. 
We present an improved algorithm 
with running time $O(n^2)$,
based on new observations on the geometric structures of the problem.
Since the input size (i.e., the total number of input visibility angles) is
$\Theta(n^2)$ in the worst case, our algorithm is worst-case optimal. 
\end{abstract}


\sectionspace
\section{Introduction}
\label{sec:intro}
\begin{figure}[t]
\begin{minipage}[t]{\linewidth}
\begin{center}
\includegraphics[totalheight=1.2in]{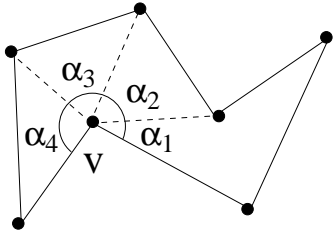}
\caption{\footnotesize Illustrating the angle measurement at a
vertex $v$: The angle sequence $(\alpha_1,\alpha_2,\alpha_3,\alpha_4)$
is given. 
}\label{fig:angles}
\end{center}
\end{minipage}
\end{figure}

\begin{figure}[t]
\begin{minipage}[t]{\linewidth}
\begin{center}
\includegraphics[totalheight=1.2in]{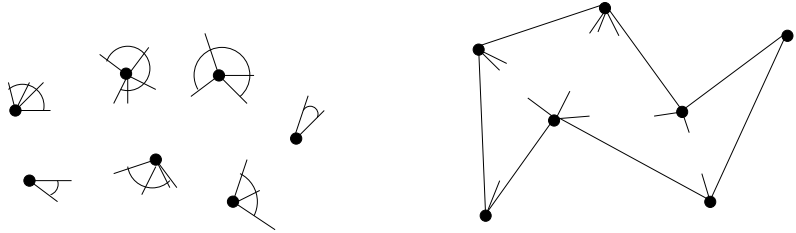}
\caption{\footnotesize The given measured angles are ordered CCW along
the polygon boundary (left); reconstructing a simple polygon that fits these angles (right).
}\label{fig:example}
\end{center}
\end{minipage}
\end{figure}

In this paper, we study the problem of reconstructing a simple polygon $P$
from the visibility angles measured at the vertices of $P$ and
from the cyclically ordered vertices of $P$ along its boundary. 
Precisely, for an unknown simple polygon
$P$ of $n$ vertices, suppose we are given (1) the vertices ordered 
counterclockwise (CCW) along the boundary 
of $P$, and (2) for each vertex $v$ of $P$, the angles
between any two adjacent rays emanating from $v$ to the vertices of
$P$ that are visible to $v$ such that these angles
are in the CCW order as seen around $v$, beginning
at the CCW neighboring vertex of $v$ on the boundary of $P$ (e.g., see
Fig.~\ref{fig:angles}). 
A vertex $u$ of $P$ is {\em visible} to a vertex $v$ of $P$ if the
line segment connecting $u$ and $v$ lies entirely in $P$.
The objective of the problem is to reconstruct the simple polygon $P$ (up
to similarity) that fits all the
given angles. We call this problem the {\em polygon
reconstruction problem from angles}, or {\em PRA} for short. 
Figure~\ref{fig:example} gives an example.


The \pra\ problem has been studied by Disser, Mihal\'ak, and Widmayer 
\cite{ref:DisserRe10}, who showed that the solution polygon 
for the input is unique (up to similarity) and gave an $O(n^3\log n)$ 
time algorithm for reconstructing such a polygon. Using the input, 
their algorithm first constructs the visibility graph $G$ of $P$ and
subsequently reconstructs the polygon $P$. As shown in 
\cite{ref:DisserRe10}, once $G$ is known, the polygon $P$ can be 
obtained efficiently (e.g., in $O(n^2)$ time) with the help of the angle data 
and the CCW ordered vertex sequence of $P$. 

Given a visibility
graph $G$, the problem of determining whether there is a polygon $P$
that has $G$ as its visibility graph is called the visibility graph
{\em recognition} problem, and the problem of actually constructing
such a polygon $P$ is called the visibility graph {\em reconstruction} problem. 
Note that the general visibility graph recognition and reconstruction
problems are long-standing open problems with only partial results known
(e.g., see \cite{ref:AsanoVi00} for a short survey). Everett
\cite{ref:EverettVi90} showed that the visibility graph reconstruction
problem is in PSPACE, but no better upper bound on the complexity of
either problem is known. In our problem setting, we have the
angle data information and the ordered vertex list of $P$; thus
$P$ can be constructed efficiently after knowing $G$.

Hence, the major part of the algorithm in \cite{ref:DisserRe10} is
dedicated to constructing the visibility graph $G$ of $P$.
As indicated in \cite{ref:DisserRe10}, the key difficulty 
is that the vertices in this
problem setting have no recognizable labels, e.g., the angle
measurement at a vertex $v$ gives angles between visible vertices to
$v$ but does not identify these visible vertices globally. 
The authors in \cite{ref:DisserRe10} also showed that some natural greedy
approaches do not seem to work. An $O(n^3\log n)$ time algorithm for
constructing $G$ is given in \cite{ref:DisserRe10}. The algorithm, 
called the {\em triangle witness algorithm}, is based
on the following observation: Suppose we wish to determine whether a
vertex $v_i$ is visible to another vertex $v_j$; then $v_i$ is
visible to $v_j$ if and only if there is a vertex $v_l$ on the portion of
the boundary of $P$ from $v_{i+1}$ to $v_{j-1}$ in the CCW order such
that $v_l$ is visible to both $v_i$ and $v_j$ and the triangle formed
by the three vertices $v_i,v_j$, and $v_l$ does not intersect the
boundary of $P$
except at these three vertices (such a vertex $v_l$ is called a {\em triangle
witness vertex}). 

In this paper, based on the triangle witness algorithm
\cite{ref:DisserRe10}, by exploiting some new geometric properties, we
give an improved algorithm with a running time of $O(n^2)$. The 
improvement is due to two key observations. First, in the triangle
witness algorithm \cite{ref:DisserRe10}, to determine whether a vertex
$v_i$ is visible to another vertex $v_j$, the algorithm needs to
determine whether there exists a triangle witness vertex along the
boundary of $P$ from $v_{i+1}$ to $v_{j-1}$ in the CCW order; to this
end, the algorithm checks every vertex in that boundary portion of 
$P$. We observe that it suffices to check only one particular vertex in that
boundary portion. This removes an $O(n)$ factor from the running time of
the triangle witness algorithm \cite{ref:DisserRe10}. Second, some
basic operations in the triangle witness algorithm
\cite{ref:DisserRe10} take $O(\log n)$ time each; by utilizing certain
different  
data structures, our new algorithm can handle each of those
basic operations in constant time. 
This removes another $O(\log n)$ factor from the running time. Note that
since the input size is $\Theta(n^2)$ in the worst case (e.g., the
total number of all visibility angles), our algorithm is worst-case optimal.  

As shown in \cite{ref:DisserRe10}, if only the angle measurements are
given, i.e., the ordered vertices along the
boundary of $P$ are unknown, then the information is not sufficient for 
reconstructing $P$. In other words, it may be possible to compute several simple
polygons that are not similar but all fit the given measured angles (see
\cite{ref:DisserRe10} for an example). 

\subsection{Related Work}

The problems of reconstructing geometric objects based on measurement
data have been studied extensively (e.g.,
\cite{ref:BiedlRe09,ref:BiloRe09,ref:JacksonOr02,ref:SidleskyPo06,ref:SnoeyinkCr99}).
As discussed above, the general visibility graph recognition and
reconstruction problems are in PSPACE \cite{ref:EverettVi90} and no better
complexity upper bound is known so far (e.g., see \cite{ref:EverettNe95}). 
Yet, some results have been given for certain special polygons. 
For example, Everett and
Corneil \cite{ref:ErerettRe90} characterized the visibility graphs of 
spiral polygons and
gave a linear time reconstruction algorithm. Choi, Shin, and Chwa
\cite{ref:ChoiCh95}, and Colley, Lubiw, and Spinrad
\cite{ref:ColleyVi97} characterized and
recognized the visibility graphs of funnel-shaped polygons. 

By adding extra information, some versions of the problems become more tractable. 
O'Rourke and Streinu \cite{ref:ORourkeTh98} considered the {\em
vertex-edge} visibility graph that includes edge-to-edge visibility
information. Wismath \cite{ref:WismathPo00} 
introduced the {\em stab graphs} which are
also an extended visibility structure and showed how parallel line
segments can be efficiently reconstructed from it. 
Snoeyink \cite{ref:SnoeyinkCr99} proved that a unique simple
polygon (up to similarity) can be determined by the interior angles at
its vertices and the cross-ratios of the diagonals of any given
triangulation, where the cross-ratio of a diagonal is the product of
the ratios of edge lengths for the two adjacent triangles. 
Jackson and Wismath \cite{ref:JacksonOr02} studied the reconstruction
of orthogonal polygons from horizontal and vertical visibility
information and gave an $O(n\log n)$ time reconstruction algorithm. 
Biedl, Durocher, and Snoeyink \cite{ref:BiedlRe09} considered the problem of
reconstructing the two-dimensional floor plan of a polygonal room
using different types of scanned data, and proposed several problem models.
Sidlesky, Barequet, and Gotsman
\cite{ref:SidleskyPo06} studied the problem of reconstructing a planar
polygon from its intersections with a collection of
arbitrarily-oriented ``cutting" lines. 

Reconstructing a simple polygon from angle data was first
considered by Bil\`o {\em et al}.~\cite{ref:BiloRe09}, who aimed to understand what kinds of
sensorial capabilities are sufficient for a robot moving inside an unknown
polygon to reconstruct the visibility graph of the polygon. 
It was shown in \cite{ref:BiloRe09} that if the robot is equipped with a compass
to measure the angle between any two vertices that are
currently visible to the robot
and also has the ability to know where it came from when
moving from vertex to vertex, 
then the visibility graph of the polygon
can be uniquely reconstructed. Reconstruction and exploration of environments 
by robots in other problem settings have also been studied (e.g.,
see \cite{ref:DudekRo91,ref:FlocchiniTh99,ref:SuriSi08}). 

The rest of this paper is organized as follows. In Section \ref{sec:pre}, we 
give the problem definitions in detail and
introduce some notations and basic observations. 
To be self-contained, in Section \ref{sec:review}, we briefly review the
triangle witness algorithm given in \cite{ref:DisserRe10}. 
We then present our improved algorithm in Section \ref{sec:improve}. 

\sectionspace
\section{Preliminaries}
\label{sec:pre}
In this section, we define the \pra\ problem in detail and introduce
the needed notations and terminology. For ease of discussion and comparison, some of our
notations follow those in \cite{ref:DisserRe10}. 

Let $P$ be a simple polygon of $n$ vertices $v_0,v_1,\ldots,v_{n-1}$ in the CCW
order along $P$'s boundary. Denote by $G=(V,E)$ the visibility graph
of $P$, where $V$ consists of all vertices of $P$ and for any two distinct
vertices $v_i$ and $v_j$, $E$ contains an edge $e(v_i,v_j)$ connecting $v_i$ and
$v_j$ if and only if $v_i$ is visible to $v_j$ inside $P$. In this
paper, the indices of all $v_i$'s are taken as congruent modulo $n$,
i.e., if $n\leq i+j\leq 2n-1$, then $v_{i+j}$ is the same vertex as $v_l$, where
$l= i+j - n$ (or $l= (i+j) \mod n$); similarly, if $-n \leq i-j<0$, then $v_{i-j}$ is the same
vertex as $v_l$, where $l=i-j+n$. For each $v_i\in V$,
denote by $deg(v_i)$ its degree in the visibility graph $G$, and denote by
$vis(v_i)=(vis_1(v_i),vis_2(v_i),\ldots,vis_{deg(v_i)}(v_i))$ 
the sequence of vertices in $P$ visible to $v_i$ from $v_{i+1}$ to
$v_{i-1}$ ordered CCW around $v_i$. We refer to $vis(v_i)$ as $v_i$'s
{\em visibility angle sequence}. 
Note that since both $v_{i-1}$ and $v_{i+1}$ are visible to $v_i$, 
$vis_1(v_{i}) = v_{i+1}$
and $vis_{deg(v_i)}(v_i) = v_{i-1}$. For any two vertices $v_i,v_j$ in
$V$, let $ch(v_i,v_j)$ denote the sequence
$(v_i,v_{i+1},\ldots,v_j)$ of the vertices ordered CCW along the
boundary of $P$ from $v_i$ to $v_j$. We refer to
$ch(v_i,v_j)$ as a {\em chain}. Let $|ch(v_i,v_j)|$ denote the
number of vertices of $P$ in the chain $ch(v_i,v_j)$. 

For any two distinct vertices $v_i,v_j \in P$, let $\rho(v_i,v_j)$ be the ray emanating from
$v_i$ and going towards $v_j$. For any three vertices $v,v_i,v_j\in P$, denote by
$\angle_v(v_i,v_j)$ the CCW angle defined by rotating
$\rho(v,v_i)$ around $v$ to $\rho(v,v_j)$ ($v_i$ or $v_j$ need not be
visible to $v$). Note that the values
of all angles we use in this paper are in $[0,2\pi)$. For any
vertex $v\in P$ and $1\leq i< j\leq deg(v)$, let $\angle_v(i,j)$ be
$\angle_v(vis_i(v),vis_j(v))$. 

The \pra\ problem can then be
re-stated as follows: Given a sequence of all vertices
$v_0,v_1,\ldots,v_{n-1}$ of an unknown simple polygon $P$ in the CCW order along
$P$'s boundary, and the angles $\angle_v(i,i+1)$ for each vertex $v$ of $P$ with
$1\leq i< deg(v)$, we seek to reconstruct $P$ (up to
similarity) to fit all the given angles. 
Without loss of generality, we assume that no three distinct vertices of $P$ are collinear.

It is easy to see that after $O(n^2)$ time preprocessing, for any $v\in V$ and any
$1\leq i< j\leq deg(v)$, the angle $\angle_v(i,j)$ can be obtained
in constant time. In the following discussion, we assume that this
preprocessing has already been done. Sometimes we (loosely) say that these
angles are given as input.

The algorithm given in \cite{ref:DisserRe10} does not construct $P$ directly.
Instead, the algorithm first computes its visibility graph $G=(V,E)$. 
As mentioned earlier, after knowing $G$, $P$ can be reconstructed efficiently with
the help of the angle data and the CCW ordered vertex sequence of $P$. 
The algorithm for constructing $G$ in \cite{ref:DisserRe10} is
called the {\em triangle witness algorithm}, which will be briefly reviewed
in Section \ref{sec:review}.
Since $V$ consists of all vertices of $P$, the
problem of constructing $G$ is equivalent to constructing its edge set
$E$, i.e., for any two distinct vertices $v_i, v_j\in P$, 
determine whether there is an edge $e(v_i,v_j)$ in $E$ connecting $v_i$
and $v_j$ (in other words,
determining whether $v_i$ is visible to $v_j$ inside $P$).

\begin{figure}[t]
\begin{minipage}[t]{\linewidth}
\begin{center}
\includegraphics[totalheight=1.5in]{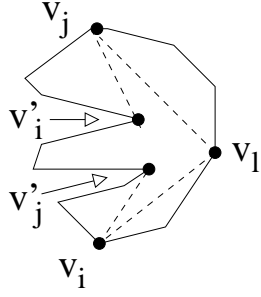}
\caption{\footnotesize Illustrating the definitions of $v'_i$ and
$v_j'$, and $\sangle_{v_i}(v_l,v_j)$ and
$\sangle_{v_j}(v_i,v_l)$ which are the angles $\angle_{v_i}(v_l,v_j')$ and
$\angle_{v_j}(v_i',v_l)$, respectively. 
}\label{fig:sangle}
\end{center}
\end{minipage}
\end{figure}

To discuss the involved algorithms, we need one more definition.
For any two vertices $v_i,v_j\in P$ with
$|ch(v_{i+1},v_{j-1})|\geq 1$, 
suppose a vertex $v_l\in ch(v_{i+1},v_{j-1})$ 
is visible to both $v_i$ and $v_j$; then we let $v_j'$ be the
{\em first} visible vertex to $v_i$ on the chain $ch(v_j,v_i)$ and let
$v_i'$ be the {\em last} visible vertex to $v_j$ on the chain $ch(v_j,v_i)$
(e.g., see Fig.~\ref{fig:sangle}). Intuitively, imagine that we
rotate a ray from $\rho(v_i,v_l)$ around $v_i$
counterclockwise; then the first vertex on the chain
$ch(v_j,v_i)$ hit by the rotating ray  is $v_j'$.
Similarly, if we rotate a ray from
$\rho(v_j,v_l)$ around $v_j$ clockwise,
then the first vertex on the chain $ch(v_j,v_i)$ hit by
the rotating ray is $v_i'$. 
Note that if $v_i$ is visible to $v_j$, then
$v_j'$ is $v_j$ and $v_i'$ is $v_i$.  We denote by 
$\sangle_{v_i}(v_l,v_j)$ the angle $\angle_{v_i}(v_l,v_j')$ and denote
by $\sangle_{v_j}(v_i,v_l)$ the angle $\angle_{v_j}(v_i',v_l)$. It should be 
pointed out that for ease of understanding this paper, the above statement of defining $\sangle$ 
is different from that in \cite{ref:DisserRe10} but they refer to the
same angles in the algorithm. The motivation for
defining $\sangle$ will be clear after discussing the
following lemma, which has been proved in \cite{ref:DisserRe10}. 


\lemmaspace
\begin{lemma}\label{lem:10}\cite{ref:DisserRe10}
For any two vertices $v_i,v_j\in P$ with $|ch(v_{i+1},v_{j-1})|\geq 1$, 
$v_i$ is visible
to $v_j$ if and only if there exists a vertex $v_l$ on
$ch(v_{i+1},v_{j-1})$ such that $v_l$ is visible to both $v_i$ and
$v_j$ and
$\sangle_{v_i}(v_l,v_j)+\sangle_{v_j}(v_i,v_l)+\angle_{v_l}(v_j,v_i)=\pi$. 
\end{lemma}
\lemmaspace

Since the above lemma is also critical to our improved algorithm in 
Section \ref{sec:improve}, we sketch the proof of the lemma below. 

For any two
vertices $v_i,v_j\in P$ with $|ch(v_{i+1},v_{j-1})|\geq 1$, if $v_i$ is visible to
$v_j$, then it is not difficult to see that 
there must exist a vertex $v_l\in ch(v_{i+1},v_{j-1})$ that is 
visible to both $v_i$ and $v_j$. Since the three vertices
$v_i,v_j$, and $v_l$ are mutually visible to each other, it is clear
that the triangle formed by these three vertices does not intersect the
boundary of $P$ except at the three vertices, implying 
that $\sangle_{v_i}(v_l,v_j)+\sangle_{v_j}(v_i,v_l)+\angle_{v_l}(v_j,v_i)=
\angle_{v_i}(v_l,v_j)+\angle_{v_j}(v_i,v_l)+\angle_{v_l}(v_j,v_i)=\pi$. 
Such a vertex $v_l$ is called a {\em triangle witness} of the edge
$e(v_i,v_j)$ in $E$. 

Suppose $v_i$ is not visible to $v_j$; then it is possible that there does not
exist a vertex $v_l\in ch(v_{i+1},v_{j-1})$ which is visible
to both $v_i$ and $v_j$. In fact, if there 
exists no vertex $v_l\in ch(v_{i+1},v_{j-1})$ that is visible
to both $v_i$ and $v_j$, then $v_i$ cannot be visible to $v_j$. Hence in
the following, we assume that such a vertex $v_l$ exists, i.e., a
vertex $v_l\in
ch(v_{i+1},v_{j-1})$ is visible to both $v_i$ and $v_j$ (yet $v_i$ is not visible to $v_j$).
Note that $v_l$ is not collinear with $v_i$ and $v_j$.
If $\angle_{v_l}(v_j,v_i)>\pi$ (i.e., the chain $ch(v_{i},v_{j})$ is part of the
boundary of $P$ that blocks
the visibility between $v_i$ and $v_j$), then the lemma obviously holds.  Otherwise,
$\angle_{v_l}(v_j,v_i)<\pi$, and in this case,
the visibility between $v_i$ and $v_j$ is blocked
by the chain $ch(v_{j},v_{i})$.
Since $v_i$ is not visible to $v_j$, for any choice of such a vertex $v_l$, 
the angle $\angle_{v_i}(v_l,v_j)$
is not given by $v_i$'s visibility angle sequence $vis(v_i)$. 
The ``closest approximation" for $\angle_{v_i}(v_l,v_j)$ in this case is determined by
a vertex $v$ on the chain $ch(v_{j},v_{i})$ such that $v$ becomes $v_j$ if and
only if $v_j$ is visible to $v_i$.
As in the definition of
$\sangle_{v_i}(v_l,v_j)$, the vertex $v_j'$ is such a vertex $v$, i.e., the
first visible vertex to $v_i$ on the chain $ch(v_j,v_i)$.
Similarly, the
vertex $v_i$ in $\angle_{v_j}(v_i,v_l)$ is ``replaced" in the definition
of $\sangle_{v_j}(v_i,v_l)$ by $v_i'$, i.e.,
the last visible vertex to $v_j$ on $ch(v_j,v_i)$.
Clearly, when $v_j$ is not visible to $v_i$, it must hold that 
$\angle_{v_i}(v_l,v_j')<\angle_{v_i}(v_l,v_j)$ and
$\angle_{v_j}(v_i',v_l)<\angle_{v_j}(v_i,v_l)$. Therefore,  
$\sangle_{v_i}(v_l,v_j)+\sangle_{v_j}(v_i,v_l)+\angle_{v_l}(v_j,v_i)=
\angle_{v_i}(v_l,v_j')+\angle_{v_j}(v_i',v_l)+\angle_{v_l}(v_j,v_i)
<\angle_{v_i}(v_l,v_j)+\angle_{v_j}(v_i,v_l)+\angle_{v_l}(v_j,v_i)=\pi$.
Lemma \ref{lem:10} thus follows. 

\sectionspace
\section{The Triangle Witness Algorithm}
\label{sec:review}
In this section, we briefly review the triangle witness algorithm in \cite{ref:DisserRe10} that
constructs the visibility graph $G=(V,E)$ of the unknown simple polygon $P$. 

The triangle witness algorithm is based on Lemma \ref{lem:10}. The
algorithm has $\lceil n/2\rceil$ iterations. In the $k$-th iteration
($1\leq k\leq \lceil n/2\rceil$),
the algorithm checks, for each $i = 0,1,\ldots, n-1$, whether $v_i$ is
visible to $v_{i+k}$. After all iterations, the edge set  
$E$ can be obtained. To this end, the algorithm maintains two maps $F$
and $B$: $F[v_i][v_j]=t$ if $v_j$ is identified as 
the $t$-th visible vertex to $v_i$ 
in the CCW order, i.e., $v_j= vis_t(v_i)$; the definition of $B$ is
the same as $F$. During the algorithm, $F$ will be filled in the CCW
order and $B$ will be filled in the clockwise (CW) order. When the algorithm 
finishes, for each $v_i$, $F[v_i]$ will have all visible vertices to $v_i$ on
the chain $ch(v_{i+1},v_{\lceil n/2\rceil})$ while $B[v_i]$ will have
all visible vertices to $v_i$ on the chain $ch(v_{\lceil
n/2\rceil}, v_{i-1})$. Thus, $F[v_i]$ and $B[v_i]$ together contain all visible
vertices of $P$ to $v_i$. 
For ease of description, we also treat $F[v_i]$ and
$B[v_i]$ as sets, e.g., $v_l\in F[v_i]$ means that there is an entry
$F[v_i][v_l]$ and $|F[v_i]|$ means the number of entries in the
current $F[v_i]$.  

Initially, when $k=1$, since every
vertex is visible to its two neighbors along the boundary of $P$, we have $F[v_i][v_{i+1}]=1$
and $B[v_i][v_{i-1}]=deg(v_i)$ for each $v_i$. In the $k$-th
iteration, we determine for each $v_i$, whether $v_i$ is visible to
$v_{i+k}$. Below, we let $j=i+k$. 
Note that $|F[v_i]|+1$ is the index of the first visible
vertex to $v_i$ in the CCW order that is not yet identified;
similarly, $(deg(v_{j})-|B[v_{j}]|)$ is the index of the first
visible vertex to $v_{j}$ in the CW order that is not yet
identified. If $v_i$ is visible to $v_{j}$, then we know that
$v_{j}$ is the $(|F[v_i]|+1)$-th visible vertex to $v_i$ and $v_i$
is the $(deg(v_{j})-|B[v_{j}]|)$-th visible vertex to $v_{j}$,
and thus we set $F[v_i][v_{j}]=|F[v_i]|+1$ and
$B[v_{j}][v_i]=deg(v_{j})-|B[v_{j}]|$. If $v_i$ is not visible to
$v_{j}$, then we do nothing. 

It remains to discuss how to determine whether $v_i$ is visible to
$v_{j}$. According to Lemma \ref{lem:10}, we need to determine
whether there exists a triangle witness vertex $v_l$ in the chain 
$ch(v_{i+1},v_{j-1})$, i.e., $v_l$ is visible to both $v_i$ and $v_{j}$ and 
$\sangle_{v_i}(v_l,v_j)+\sangle_{v_j}(v_i,v_l)+\angle_{v_l}(v_j,v_i)=\pi$. 
To this end, the algorithm checks every vertex $v_l$ in
$ch(v_{i+1},v_{j-1})$. For each $v_l$, the algorithm first determines 
whether $v_l$ is visible to both
$v_i$ and $v_{j}$, by checking whether there is an entry for $v_l$ in $F[v_i]$
and in $B[v_{j}]$. The algorithm utilizes balanced binary search trees
to represent $F$ and $B$, and thus checking whether there is an entry
for $v_l$ in $F[v_i]$ and in $B[j]$ can be done in $O(\log n)$ time. If $v_l$ is
visible to both $v_i$ and $v_{j}$, then the next step is to determine whether
$\sangle_{v_i}(v_l,v_j)+\sangle_{v_j}(v_i,v_l)+\angle_{v_l}(v_j,v_i)=\pi$. 
It is easy to know that $\angle_{v_l}(v_j,v_i)$ is 
$\angle_{v_l}(F[v_l][v_{j}],B[v_l][v_i])$, which can be found readily from the input.
To obtain $\sangle_{v_i}(v_l,v_j)$, we claim that it is the angle
$\angle_{v_i}(F[v_i][v_l],|F[v_i]|+1)$. Indeed, observe that all visible
vertices in the chain $ch(v_{i+1},v_{j-1})$ are in the current
$F[v_i]$. As explained above, $|F[v_i]|+1$ is the index of the first visible  
vertex to $v_i$ in the CCW order that has not yet been identified, which is
the first visible vertex to $v_i$ in the chain $ch(v_j,v_i)$, i.e, the
vertex $v_j'$. Thus, $\angle_{v_i}(F[v_i][v_l],|F[v_i]|+1)$ is
$\angle_{v_i}(v_l,v_j')$, which is $\sangle_{v_i}(v_l,v_j)$.
Similarly, $\sangle_{v_j}(v_i,v_l)$ is the angle
$\angle_{v_j}(deg(v_j)-|B[v_j]|,B[v_j][v_l])$. Note that both the angles
$\angle_{v_i}(F[v_i][v_l],|F[v_i]|+1)$ and
$\angle_{v_j}(deg(v_j)-|B[v_j]|,B[v_j][v_l])$
are known. Algorithm \ref{algo:10} in the Appendix summarizes the whole 
algorithm \cite{ref:DisserRe10}.

To analyze the running time of the above triangle witness algorithm,
note that it has $\lceil n/2\rceil$ iterations. In each iteration, the
algorithm checks whether $v_i$ is visible to $v_{i+k}$ for
each $0\leq i\leq n-1$. For each $v_i$, the algorithm checks every $v_l$
for $i+1\leq l\leq i+k-1$, i.e, in the chain $ch(v_{i+1},v_{i+k-1})$. 
For each such $v_l$, the algorithm takes $O(\log n)$ time as it 
uses balanced binary search trees to represent the two maps $F$ and $B$. 
In summary, the overall running time of the triangle witness algorithm in
\cite{ref:DisserRe10} is $O(n^3\log n)$. 

\sectionspace
\section{An Improved Triangle Witness Algorithm}
\label{sec:improve}

In this section, we present an improved solution over the triangle
witness algorithm in \cite{ref:DisserRe10} 
sketched in Section \ref{sec:review}. Our improved
algorithm runs in $O(n^2)$ time. Since the input size (e.g., the total
number of visibility angles) is $\Theta(n^2)$ in the worst case, our improved
algorithm is worst-case optimal. Our new algorithm follows the high-level 
scheme of the triangle witness algorithm in \cite{ref:DisserRe10}, and thus
we call it the {\em improved triangle witness algorithm}. 

As in \cite{ref:DisserRe10}, the new algorithm also has $\lceil n/2\rceil$
iterations, and in each iteration, we determine 
whether $v_i$ is visible to $v_{i+k}$ for each $0\leq i\leq n-1$.
For every pair of vertices $v_i$ and $v_{i+k}$, let $j=i+k$. To
determine whether $v_i$
is visible to $v_j$, the triangle witness algorithm \cite{ref:DisserRe10}
checks each vertex $v_l$ in the chain $ch(v_{i+1},v_{j-1})$ to 
see whether there exists a triangle witness vertex. 
In our new algorithm, instead, we claim that we 
need to check only one particular vertex, $vis_{|F[v_i]|}(v_i)$, i.e., the
last visible vertex to $v_i$ in the chain $ch(v_{i+1},v_{j-1})$ in the
CCW order, as stated in the following lemma. 

\begin{figure}[t]
\begin{minipage}[t]{\linewidth}
\begin{center}
\includegraphics[totalheight=1.5in]{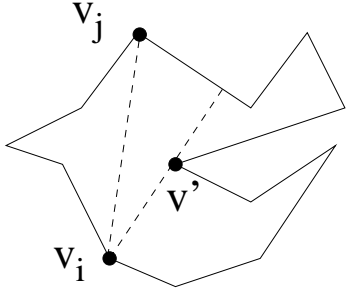}
\caption{\footnotesize Illustrating Lemma \ref{lem:20}: $v'$ is
$vis_{|F[v_i]|}(v_i)$, the
last visible vertex to $v_i$ in $ch(v_{i+1},v_{j-1})$; $v_i$ is
visible to $v_j$ if and only if $v'$ is a triangle witness vertex.
}\label{fig:visible}
\end{center}
\end{minipage}
\end{figure}

\begin{figure}[t]
\begin{minipage}[t]{\linewidth}
\begin{center}
\includegraphics[totalheight=1.5in]{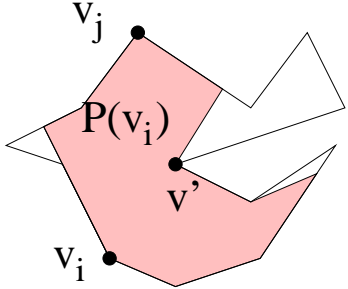}
\caption{\footnotesize Illustrating the visibility polygon $P(i)$ 
for the example in Fig.~\ref{fig:visible}: The gray (pink) area is
$P(i)$, which is a star-shaped polygon with the vertex $v_i$ as a kernel point. 
}\label{fig:visPolygon}
\end{center}
\end{minipage}
\end{figure}

\begin{figure}[t]
\begin{minipage}[t]{\linewidth}
\begin{center}
\includegraphics[totalheight=1.5in]{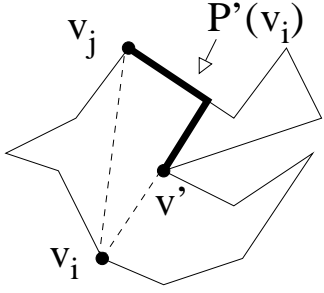}
\caption{\footnotesize Illustrating $P'(i)$ for the example in
Fig.~\ref{fig:visible}: The thick line segments form $P'(i)$. 
}\label{fig:boundary}
\end{center}
\end{minipage}
\end{figure}

\lemmaspace
\begin{lemma}\label{lem:20}
The vertex $v_i$ is visible to $v_j$ if and only if the vertex
$vis_{|F[v_i]|}(v_i)$ is a triangle witness vertex for $v_i$ and $v_j$. 
\end{lemma}
\lemmaspace
\begin{proof}
Let $v'$ denote the vertex $vis_{|F[v_i]|}(v_i)$, which is the last visible vertex
to $v_i$ in the chain $ch(v_{i+1},v_{j-1})$ in the CCW order.
Recall that $v'$ being a triangle witness vertex for $v_i$ and $v_j$
is equivalent to saying that $v'$ is visible to both $v_i$ and $v_j$ and 
$\sangle_{v_i}(v',v_j)+\sangle_{v_j}(v_i,v')+\angle_{v'}(v_j,v_i)=\pi$. 

If $v_i$ is visible to $v_j$, then we prove below that $v'$ is a triangle
witness vertex for $v_i$ and $v_j$, i.e., we prove that $v'$ is visible to both $v_i$ and
$v_j$ and
$\sangle_{v_i}(v',v_j)+\sangle_{v_j}(v_i,v')+\angle_{v'}(v_j,v_i)=\pi$. 
Refer to Fig.~\ref{fig:visible} for an example. 
Let $P(v_i)$ denote the subpolygon of $P$ that is visible to
the vertex $v_i$. Usually, $P(v_i)$ is called the {\em visibility
polygon} of $v_i$ and it is well-known that $P(v_i)$ is
a star-shaped polygon with $v_i$ as a kernel point
(e.g., see \cite{ref:AsanoVi00}). Figure~\ref{fig:visPolygon}
illustrates $P(v_i)$.
Since $v'$ is the last visible vertex to $v_i$ in
$ch(v_{i+1},v_{j-1})$ and $v_j$ is visible to $v_i$, we claim that
$v_j$ is also visible to $v'$. Indeed, since both $v'$ and $v_j$ are visible to $v_i$, $v'$
and $v_j$ are both on the boundary of the visibility polygon $P(v_i)$.  
Let $P'(v_i)$ denote the portion of the boundary of $P(v_i)$ from $v'$ to
$v_j$ counterclockwise (see Fig.~\ref{fig:boundary}). 
We prove below that $P'(v_i)$ does not contain
any vertex of $P$ except $v'$ and $v_j$.
First, $P'(v_i)$ cannot contain any vertex in the chain
$ch(v_{j+1},v_{i-1})$ (otherwise, $v_j$ would not be visible to
$v_i$). Let the index of the vertex $v'$ be $l$, i.e.,
$v'= v_l$. Similarly, $P'(v_i)$ cannot contain any vertex in the chain
$ch(v_{i+1},v_{l-1})$ (otherwise, $v'$ would not be visible to
$v_i$). Finally, $P'(v_i)$ cannot contain any vertex in the chain
$ch(v_{l+1},v_{j-1})$, since otherwise $v'$ would not be the last
visible vertex to $v_i$ in $ch(v_{i+1},v_{j-1})$. Thus, $P'(v_i)$ does
not contain any vertex of $P$ except $v'$ and $v_j$. Therefore, the
region bounded by $P'(v_i)$, the line segment connecting $v_i$ and
$v_j$, and the line segment connecting $v_i$ and $v'$ must be convex (in fact,
it is always a triangle) and this region is entirely contained in $P$.  This
implies that $v_j$ is visible to $v'$. 
Hence, the three vertices $v_i,v_j$, and $v'$ are mutually visible to each
other, and we have 
$\sangle_{v_i}(v_l,v_j)+\sangle_{v_j}(v_i,v_l)+\angle_{v_l}(v_j,v_i)=
\angle_{v_i}(v_l,v_j)+\angle_{v_j}(v_i,v_l)+\angle_{v_l}(v_j,v_i)=\pi$. 

On the other hand, if $v'$ is a triangle witness vertex 
for $v_i$ and $v_j$, then by Lemma
\ref{lem:10}, the vertex $v_i$ is visible to $v_j$. The lemma thus
follows.  
\end{proof}

By Lemma \ref{lem:20}, to determine whether $v_i$ is visible to
$v_j$, instead of checking every vertex in $ch(v_{i+1},v_{j-1})$, we
need to consider only the vertex $vis_{|F[v_i]|}(v_i)$ in the current set $F[v_i]$. 
Hence, Lemma \ref{lem:20} immediately reduces the running time of the triangle
witness algorithm by an $O(n)$ factor. The other $O(\log n)$ factor
improvement is due to a new way of defining and representing the maps $F$ and
$B$, as elaborated below. In the following discussion.
let $v'$ be the vertex $vis_{|F[v_i]|}(v_i)$. 

In our new algorithm, to determine whether $v_i$ is visible
to $v_j$, we check whether $v'$ is a triangle witness vertex for $v_i$ and $v_j$.
To this end, we already
know that $v'$ is visible to $v_i$, but we still need to check whether $v'$
is visible to $v_j$. In the previous triangle witness algorithm
\cite{ref:DisserRe10}, this
step is performed in $O(\log n)$ time by representing $F$ and $B$ using
balanced binary search trees. In our new algorithm, we handle this
step in $O(1)$ time, by redefining $F$ and $B$ and using a new way to
represent them. 

We redefine $F$ as follows: $F[v_i][v_j]=t$
if $v_j$ is the $t$-th visible vertex to $v_i$ in the CCW order; if
$v_j$ is not visible to $v_i$ or $v_j$ has not yet been identified, 
then $F[v_i][v_j]=0$. For convenience, we let
$F[v_i]=F[v_i][v_{i+1},v_{i+2},\ldots,v_{i-1}]$. 
Thus, in our new definition, the size of
$F[v_i]$ is fixed throughout the algorithm, i.e., $|F[v_i]|$ is always
$n-1$. In
addition, for each $v_i$, the new algorithm maintains two variables
$L_i$ and $I_i$ for $F[v_i]$, where $L_i$ is the number of non-zero
entries in the current $F[v_i]$, which is also the number of visible
vertices to $v_i$ that have been identified 
(i.e., the number of visible vertices to $v_i$ 
in the chain $ch(v_{i+1},v_{i+k-1})$) up to the $k$-th iteration, 
and $I_i$ is the index  
of the last non-zero entry in the current $F[v_i]$, i.e., $I_i$ is the
last visible vertex to $v_i$ in the chain $ch(v_{i+1},v_{i+k-1})$ in
the CCW order. 
Similarly, we redefine $B$ in the same way as $F$, i.e., for
each $v_i$, $B[v_i]=B[v_i][v_{i+1},v_{i+2},\ldots,v_{i-1}]$ and $B[v_i][v_j]=t$
if $v_j$ is the $t$-th visible vertex to $v_i$ in the CCW order.  
Further, for each $v_i$, we also maintain 
two variables $L_i'$ and $I_i'$ for $B[v_i]$, where $L_i'$ is the
number of non-zero entries in the current $B[v_i]$, which is also the
number of visible vertices to $v_i$ in the chain
$ch(v_{i-k+1},v_{i-1})$ (up to the $k$-th iteration), and $I_i'$ is the
index of the first non-zero entry in the current $B[v_i]$, i.e., $I_i'$ 
is the first visible vertex to $v_i$ in the chain
$ch(v_{i-k+1},v_{i-1})$ in the CCW order. During the
algorithm, the array $F[v_i]$ will be filled in the CCW order, i.e., 
from the first entry 
$F[v_i][v_{i+1}]$ to the end while the array $B[v_i]$ will be filled
in the CW order, i.e, from the last entry $B[v_i][v_{i-1}]$ to the
beginning. When the algorithm finishes, $F[v_i]$ will 
contain all the visible vertices to $v_i$ in the chain
$ch(v_{i+1},v_{i+\lceil n/2 \rceil})$, and thus only the entries of the first half
of $F[v_i]$ are possibly filled with non-zero values.
Similarly, only the entries of the second half of $B[v_i]$ are possibly
filled with non-zero values. Below, we discuss the implementation details of
our new algorithm, which is summarized in Algorithm \ref{algo:20}.

\begin{algorithm}[t]
\caption{The Improved Triangle Witness Algorithm}
\label{algo:20}
\KwIn{$n$, $(v_1,v_2,\ldots,v_n)$, $deg(v_i)$, and the angles
$\angle_{v_i}(\cdot,\cdot)$ for each $0\leq i\leq n-1$.}
\KwOut{The edge set $E$ of the visibility graph $G$ for the target polygon $P$.}
\BlankLine
\tcc{All indices below are understood modulo $n$.}
$E\leftarrow \emptyset$\;
\For{$i\leftarrow 0$ \KwTo $n-1$}{
$F[v_i][v_{i+1},v_{i+2},\ldots,v_{i-1}]\leftarrow 0$\;
$B[v_i][v_{i+1},v_{i+2},\ldots,v_{i-1}]\leftarrow 0$\;
}
\For{$i\leftarrow 0$ \KwTo $n-1$}{
$E\leftarrow E\cup \{e(v_i,v_{i+1})\}$\;
$F[v_i][v_{i+1}]\leftarrow 1$\;
$B[v_i][v_{i-1}]\leftarrow deg(v_{i})$\;
$L_i\leftarrow 1, I_i\leftarrow v_{i+1}, L'_i\leftarrow 1,
I_i'\leftarrow v_{i-1}$\;
}

\For{$k\leftarrow 2$ \KwTo $\lceil n/2 \rceil$}{
\For{$i\leftarrow 0$ \KwTo $n-1$}{
$j\leftarrow i+k$\;
$v'\leftarrow I_i$\;
\If{$B[v_j][v']>0$}{
$\sangle_{v_i}(v',v_j)\leftarrow \angle_{v_i}(F[v_i][v'],L_i+1)$\;
$\sangle_{v_j}(v_i,v')\leftarrow \angle_{v_j}(deg(v_j)-L_i',B[v_j][v'])$\;
$\angle_{v'}(v_j,v_i)\leftarrow \angle_{v'}(F[v'][v_j],B[v'][v_i])$\;
}
\If{$\sangle_{v_i}(v',v_j)+\sangle_{v_j}(v_i,v')+\angle_{v'}(v_j,v_i)=\pi$}{
$E\leftarrow E\cup \{e(v_i,v_j)\}$\;
$F[v_i][v_j]\leftarrow L_i+1$\;
$B[v_j][v_i]\leftarrow deg(v_j)-L_j'$\;
$L_i\leftarrow L_i+1$, $L'_j\leftarrow L'_j+1$, $I_i\leftarrow v_j$,
$I'_j\leftarrow v_i$\;
}
}
}
\end{algorithm}

Initially, when $k=1$, for each $v_i$, we set $F[v_i][v_{i+1}]=1$ and
$B[v_i][v_{i-1}]=deg(v_i)$, and set all other entries
of $F[v_i]$ and $B[v_i]$ to zero. In addition, we set
$L_i=1$, $I_i=v_{i+1}$ and $L'_i=1$, $I'_i=v_{i-1}$.  In the $k$-th
iteration, with $1<k\leq \lceil n/2 \rceil$, for each $v_i$, 
we check whether $v_i$ is visible to $v_j$,
with $j=i+k$. If $v_i$ is not visible to $v_j$, then we do nothing.
Otherwise, we set $F[v_i][v_j]=L_i+1$ and increase $L_i$ by one;
similarly, we set $B[v_j][v_i]=deg(v_j)-L'_j$ and increase $L'_j$ by one.
Further, we set $I_i=v_j$ and $I_j'=v_i$. 

It remains to show how to check whether $v_i$ is visible to $v_j$. 
By Lemma \ref{lem:20}, we need to determine whether
$v'$ is a triangle witness vertex for $v_i$ and $v_j$. Since $v'$ is the last visible
vertex to $v_i$ in the chain $ch(v_{i+1},v_{j-1})$ in the CCW order, 
based on our definition, $v'$ is the vertex $I_i$. After knowing
$v'$, we then check whether $v'$ is visible to $v_j$, which can be
done by checking whether $B[v_j][v']$ is zero, in constant time. 
If $B[v_j][v']$ is
zero, then $v'$ is not visible to $v_j$ and $v'$ is not a triangle
witness vertex; otherwise, $v'$ is visible to
$v_j$. (Note that we can also check whether $F[v'][v_j]$ is zero.)
In the following, we assume that $v'$ is visible to $v_j$. The next step is
to determine whether  
$\sangle_{v_i}(v',v_j)+\sangle_{v_j}(v_i,v')+\angle_{v'}(v_j,v_i)=\pi$.
To this end, we must know the involved three angles. Similar
to the discussion in Section \ref{sec:review}, we have 
$\angle_{v'}(v_j,v_i)=\angle_{v'}(F[v'][v_j],B[v'][v_i])$, 
$\sangle_{v_i}(v',v_j)=\angle_{v_i}(F[v_i][v'],L_i+1)$, and
$\sangle_{v_j}(v_i,v')=\angle_{v_j}(deg(v_j)-L_j',B[v_j][v'])$. 
Thus, all these three angles can be
obtained in constant time. Hence, the step of checking whether $v_i$
is visible to $v_j$ can be performed in constant time, which reduces
another $O(\log n)$ factor from the running time of the previous
triangle witness algorithm in \cite{ref:DisserRe10}.  
Algorithm \ref{algo:20} summarizes the whole algorithm. Clearly, the running
time of our new algorithm is bounded by $O(n^2)$. 

\lemmaspace
\begin{theorem}\label{theo:10}
Given the visibility angles and an ordered vertex sequence of a simple
polygon $P$, the improved triangle witness algorithm can reconstruct
$P$ (up to similarity) in $O(n^2)$ time. 
\end{theorem}
\lemmaspace

As discussed in \cite{ref:DisserRe10}, the above algorithm can also be used
to determine whether the input angle data are consistent. Namely, if
there is no polygon that can fit the input angle data, then 
the algorithm in Theorem \ref{theo:10} can be used to report this
case as well.



\newpage
\normalsize
\appendix
\section*{Appendix}

\begin{algorithm}
\caption{The Triangle Witness Algorithm \cite{ref:DisserRe10}}
\label{algo:10}
\KwIn{$n$, $(v_1,v_2,\ldots,v_n)$, $deg(v_i)$, and the angles
$\angle_{v_i}(\cdot,\cdot)$ for each $0\leq i\leq n-1$.}
\KwOut{The edge set $E$ of the visibility graph $G$ for the target polygon $P$.}
\BlankLine
\tcc{All indices below are understood modulo $n$.}
$F\leftarrow [\text{array of n empty maps}]$\; 
$B\leftarrow [\text{array of n empty maps}]$\;
$E\leftarrow \emptyset$\;
\For{$i\leftarrow 0$ \KwTo $n-1$}{
$E\leftarrow E\cup \{e(v_i,v_{i+1})\}$\;
$F[v_i][v_{i+1}]\leftarrow 1$\;
$B[v_{i}][v_{i-1}]\leftarrow deg(v_{i})$\;
}
\For{$k\leftarrow 2$ \KwTo $\lceil n/2 \rceil$}{
\For{$i\leftarrow 0$ \KwTo $n-1$}{
$j\leftarrow i+k$\;
\For{$l\leftarrow i+1$ \KwTo $j-1$}{
\If{$v_l\in F[v_i]$ and $v_l\in B[v_j]$}{
$\sangle_{v_i}(v_l,v_j)\leftarrow \angle_{v_i}(F[v_i][v_l],|F[v_i]|+1)$\;
$\sangle_{v_j}(v_i,v_l)\leftarrow
\angle_{v_j}(deg(v_j)-|B[v_j]|,B[v_j][v_l])$\;
$\angle_{v_l}(v_j,v_i)\leftarrow
\angle_{v_l}(F[v_l][v_j],B[v_l][v_i])$\;
}
\If{$\sangle_{v_i}(v_l,v_j)+\sangle_{v_j}(v_i,v_l)+\angle_{v_l}(v_j,v_i)=\pi$}{
$E\leftarrow E\cup \{e(v_i,v_j)\}$\;
$F[v_i][v_j]\leftarrow |F[v_i]|+1$\;
$B[v_j][v_i]\leftarrow deg(v_j)-|B[v_j]|$\;
abort the innermost loop\;
}
}
}
}
\end{algorithm}

\end{document}